\documentclass{article} 

\usepackage{spconf}
\usepackage[utf8]{inputenc}
\usepackage{CJKutf8}
\usepackage{tabularx}
\usepackage{array}
\usepackage{amssymb}
\usepackage{amsmath}
\usepackage{graphicx}
\usepackage{comment}

\usepackage{smartdiagram}
\usepackage{tikz}
\usetikzlibrary{plotmarks}
\usetikzlibrary{math}

\title{Improved Robust ASR for Social Robots in Public Spaces}

\author{Charles Jankowski, Vishwas Mruthyunjaya, Ruixi Lin, and Ashley Zeng\\CloudMinds Technology Inc\\Santa Clara, CA, USA\\ charles.jankowski@cloudminds.com}

\name{Charles Jankowski, Vishwas Mruthyunjaya, and Ruixi Lin}

\address{Cloudminds Technology Inc.\\4500 Great America Parkway\\Santa Clara, CA, USA}

\begin{document}

\maketitle

\begin{abstract}

Social robots deployed in public spaces present a challenging task for ASR because of a variety of factors, including noise SNR of 20 to 5 dB. Existing ASR models perform well for higher SNRs in this range, but degrade considerably with more noise. This work explores methods for providing improved ASR performance in such conditions. We use the AiShell-1 Chinese speech corpus and the Kaldi ASR toolkit for evaluations. We were able to exceed state-of-the-art ASR performance with SNR lower than 20 dB, demonstrating the feasibility of achieving relatively high performing ASR with open-source toolkits and hundreds of hours of training data, which is commonly available.
\end{abstract}



\section{PROBLEM DESCRIPTION}

The increased adoption of social robots deployed in very public spaces such as malls, stores, and airports, has made the requirements for high-performance speech recognition significantly higher.

\subsubsection{Social Robots}

\begin{figure}[thpb]
      \centering
      \framebox{\parbox{3in}{\includegraphics[scale=0.15]{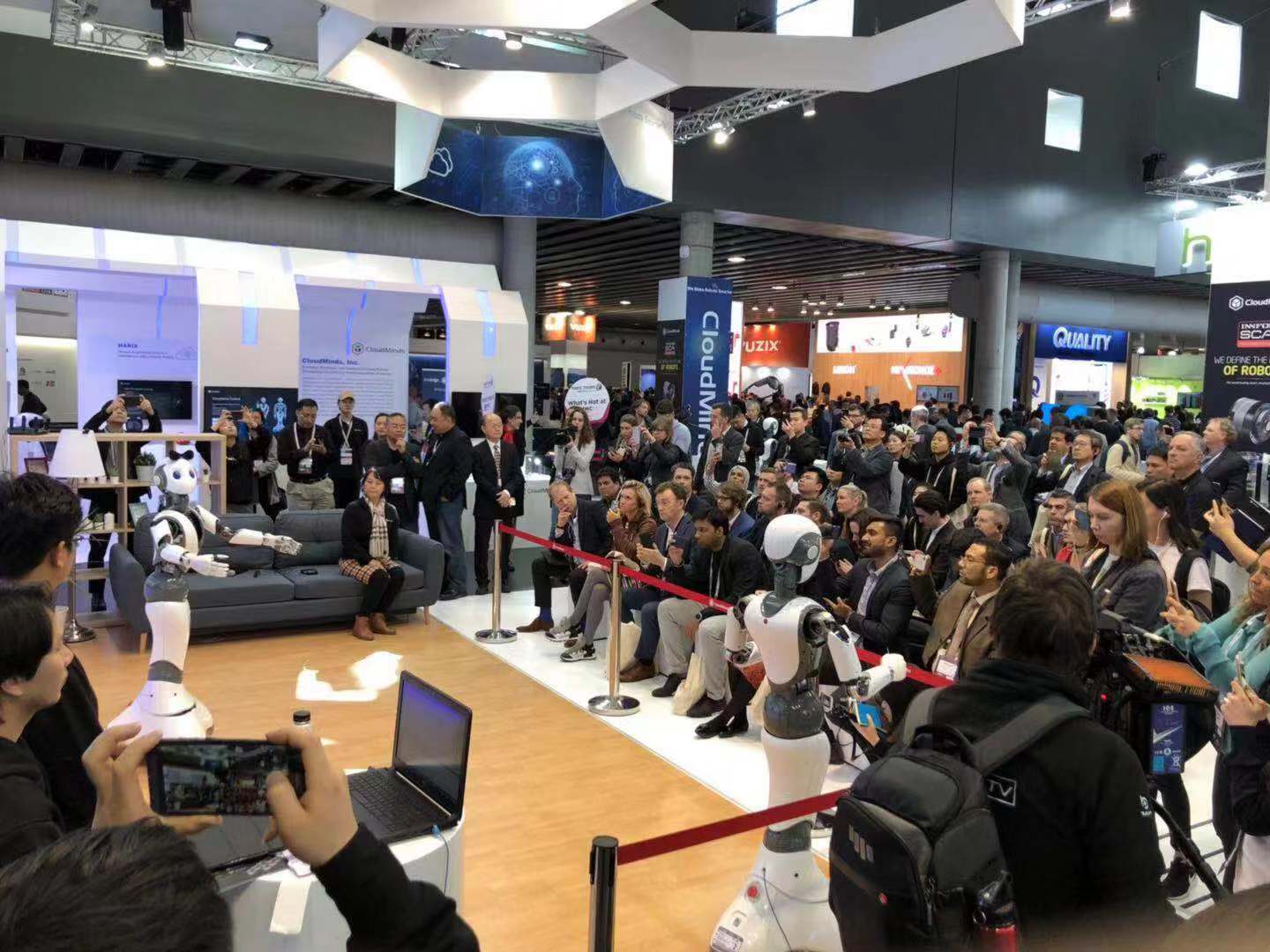}}}
      \caption{CloudMinds' XR-1 robot.}
      \label{XR1}
   \end{figure}

\begin{figure}[thpb]
      \centering
      \framebox{\parbox{3in}{\includegraphics[scale=0.205]{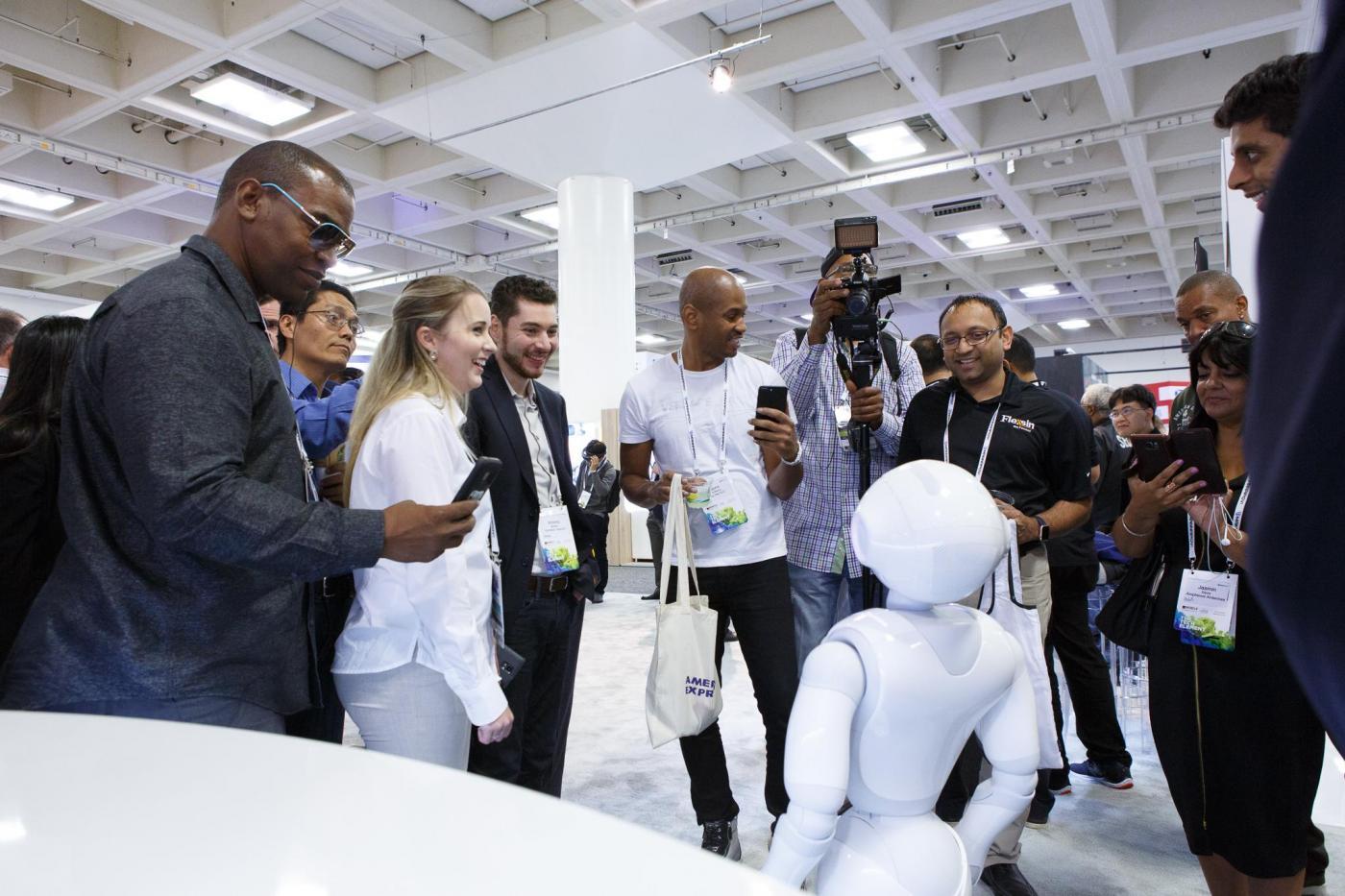}}}
      \caption{The Pepper social robot.}
      \label{PepperCrowd}
   \end{figure}

\begin{figure}[thpb]
      \centering
      \framebox{\parbox{3in}{\includegraphics[scale=0.17]{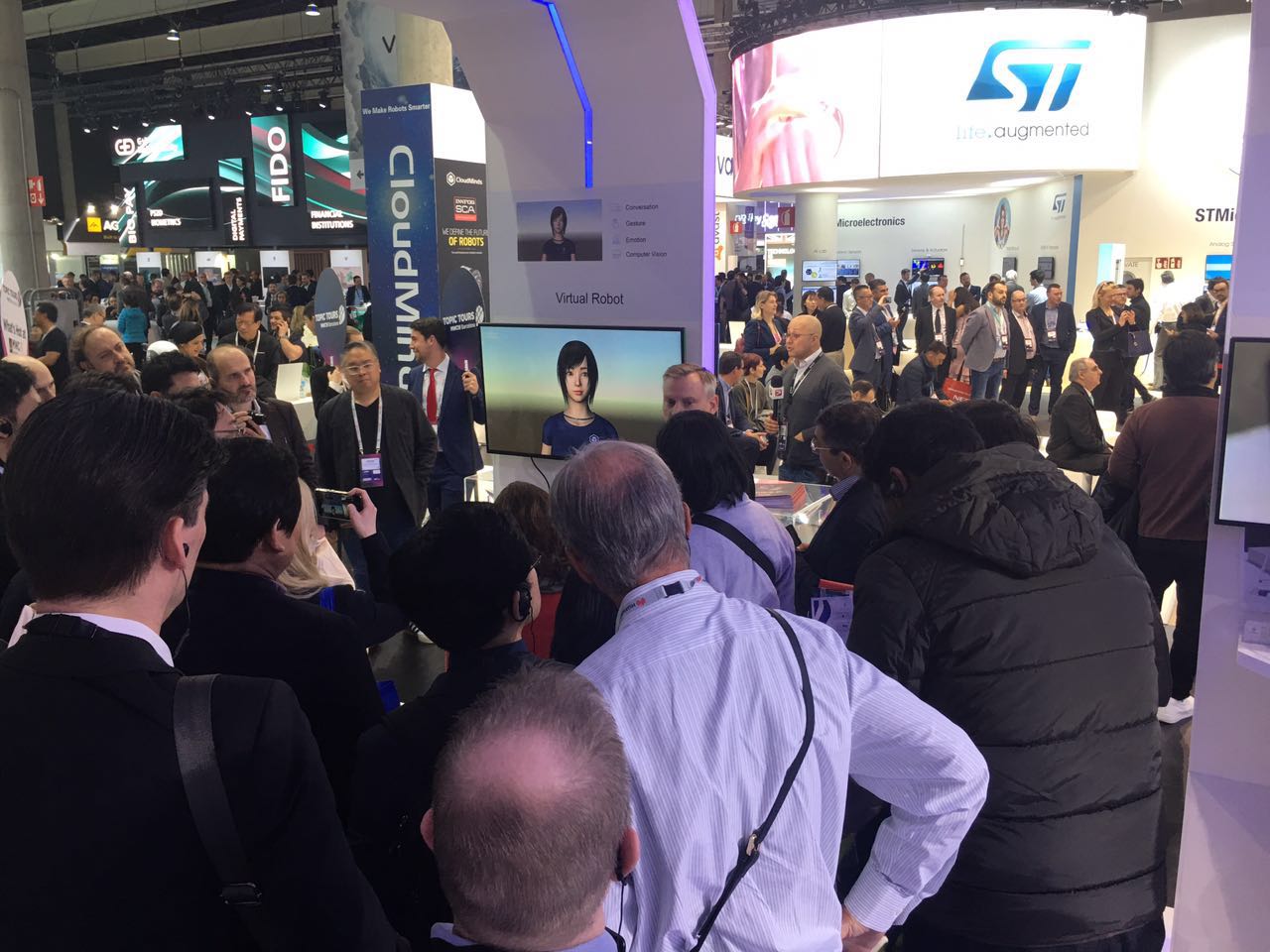}}}
      \caption{The Cloudia intelligent digital avatar from CloudMinds.}
      \label{Cloudia}
   \end{figure}

Here we show examples of various social robots. Figure \ref{XR1} shows CloudMinds' XR-1 robot. Figure \ref{PepperCrowd} shows Pepper, a social robot made by SoftBank Robotics. Of a slightly different nature is the Cloudia intelligent digital avatar from CloudMinds, shown in Figure \ref{Cloudia}.

In contrast to some other social robots that are deployed more often in settings such as homes, Pepper is primarily geared toward installations in public spaces. The very public nature of Pepper and similar robots, as well as the sites and environments in which they are operating, have a large bearing on the requirements for various components of the robot such as ASR.
   
\subsubsection{Motivation for Physical Robots and Avatars}

With the proliferation of text bots on mobile phones and elsewhere, as well as "voice bots" such as Amazon Echo and Google Home, it is worth considering the advantages of intelligent digital avatars and physical social robots.

\begin{figure}[thpb]
      \centering
      \framebox{\parbox{3in}{\includegraphics[scale=0.3]{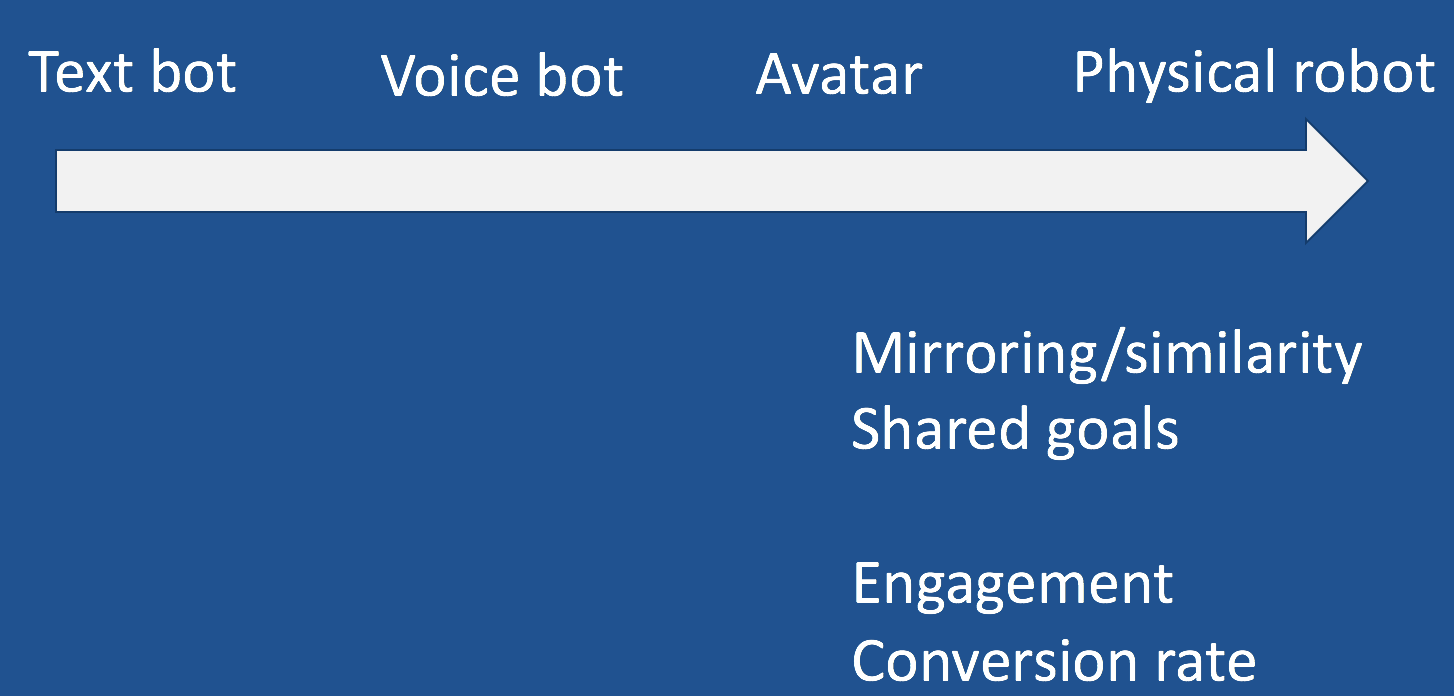}}}
      \caption{Robotic form factors and engagement.}
      \label{Engagement}
   \end{figure}

We have found \cite{FTC2019} through deploying physical social robots and avatars, that with devices and/or environments that more closely resemble the form of human subjects (either physically or virtually), we can achieve a higher level of user engagement than with other form factors of conversational systems. At this time, we do not have metrics to quantify the concept of use "engagement", but more informal observation. Figure \ref{Engagement} shows this.

This potential for increased user engagement motivates the desire to address the significantly challenging conditions where these robots are deployed.

\section{Conditions for ASR with Public Social Robots}

Here we consider under what conditions these robots are deployed from an ASR perspective, and what major challenges exist.

\subsection{Factors Dictating ASR Performance}

There are several factors the influence ASR performance in challenging conditions. Performance in this case is typically quantified by word error rate (WER) or character error rate (CER) for Asian languages. These factors include:
\begin{itemize}
    \item Vocabulary: It is well-known that the perplexity of the language model has a significant inverse effect on performance.
    \item Microphone distance: Speakers further away from the microphone, especially in acoustically active rooms, can result in substantially lower performance.
    \item Noise: One of the biggest factors affecting ASR performance is the noise level, typically quantified as signal-to-noise ratio (SNR).
    \item Reverberation: Highly reverberant acoustic environments are particularly challenging. Reverberation time is typically quantified by $T_{60}$, or the time in seconds for sounds in a room to decay by 60 dB. Table \ref{T60} shows $T_{60}$ for various acoustic environments.
\end{itemize}

\begin{table}
    \centering
    \begin{tabular}{|l|l|}
    \hline
    Environment & $T_{60}$ \\
    \hline
    Home & 0.55 \\
    Office & 0.4--0.7 \\
    Mall & 1.7--3.2 \\
    Airport & 3+ \\
    \hline
    \end{tabular}
    \caption{$T_{60}$ for various acoustic environments.}
    \label{T60}
\end{table}

\subsubsection{Noise for Robot Deployments}
\label{robotnoise}

\begin{figure}[thpb]
      \centering
      \begin{tikzpicture}[y=.06cm, x=.22cm,font=\sffamily]
	\draw (-10,0) -- coordinate (x axis mid) (25,0);
    \foreach \x in {25,20,15,10,5,0,-5,-10}
    	\tikzmath{
        	\y = 25 - \x;
        }
     	\draw (\x,1pt) -- (\x,-3pt)
			node[anchor=north] {\y};
    
	\node[below=0.8cm] at (x axis mid) {SNR (dB)};

	\draw (5,40) rectangle (20,50) node[pos=0.5] {Robot SNR Range};
    
	\begin{scope}[shift={(-10,15)}] 
    \fill (-1,2) -- (1,2) -- (0, -2) -- (-1,2) node[right=10pt] {AiShell-1};
    \end{scope}
    
    \begin{scope}[shift={(5,30)}] 
    \fill (-1,2) -- (1,2) -- (0, -2) -- (-1,2) node[right=10pt] {Google Home};
    \end{scope}
    
	\draw (5,18) rectangle (15,25) node[pos=0.5] {Home, inside};
   
    \begin{scope}[shift={(15,15)}] 
    \fill (-1,2) -- (1,2) -- (0, -2) -- (-1,2) node[right=10pt] {Hospital room};
    \end{scope}
    
    \begin{scope}[shift={(20,8)}] 
    \fill (-1,2) -- (1,2) -- (0, -2) -- (-1,2) node[left] {Hospital, public space};
    \end{scope}
    
\end{tikzpicture}
      \caption{SNR of various conditions, including public deployments of social robots, and the AiShell-1 data.}
      \label{SNRfigure}
   \end{figure}
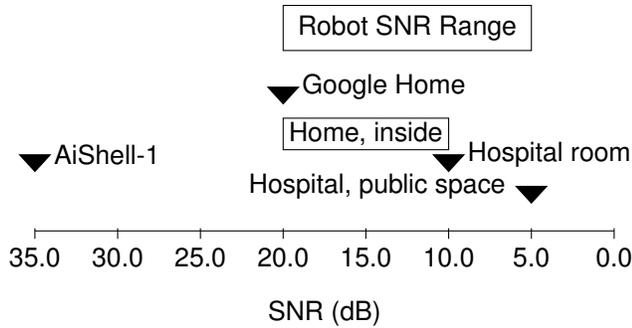

CloudMinds has a number of social robots and avatars such as XR-1, Pepper, and Cloudia (shown in figures \ref{XR1}, \ref{PepperCrowd}, and \ref{Cloudia}) deployed in various public spaces such as hospitals, stores, malls, and office buildings. We recorded raw speech from some of these deployments and measured the noise conditions, notably signal-to-noise-ratio (SNR) measured similarly to \footnote{https://www.nist.gov/information-technology-laboratory/iad/mig/nist-speech-signal-noise-ratio-measurements}, and compared to SNRs from other well-known conditions and environments \cite{SNRenvironment}. Figure \ref{SNRfigure} shows a comparison. 

We see from figure \ref{SNRfigure} that public deployments of robots normally operate from 15--20 dB SNR for a relatively quiet office, to 5--7 dB for a loud trade show in a very reverberant environment. This is in contrast, for example, to home-based social robots such as Alexa or Google Home, which experience an SNR of about 20 dB \cite{GoogleHomeSNR}. We will see later that many ASR systems perform very well in clean or 20 dB SNR speech, but start degrading past 20 dB, and show quite substantial errors at 10 dB SNR and beyond.

Table \ref{ASRFactors} summarizes some of these factors that contribute to ASR performance, for the popular Alexa/Google Home voice bot devices as well as several of the research challenges such as CHiME \cite{chime-1,chime-2,chime-3,chime-4,chime-5} and REVERB \cite{reverb}, and finally what we have observed from deploying physical social robots, as described in section \ref{robotnoise}.

\begin{table*}[t]
    \centering
    \begin{tabular}{|p{80mm}|l|p{15mm}|l|l|p{15mm}|}
    \hline
    Use case/challenge & Perplexity & Microphone distance (m) & SNR (dB) & $T_{60}$ & WER \\
    \hline
    \hline
    Alexa/Google Home (2016) & Medium & 2 & 20 & 0.5 & 20 \\
    \hline
    CHiME-1: Binaural living room (2011) \cite{chime-1}& 6.3 & 2 & 9 to -6 & 0.3 & 8.2@6 dB \\
    CHiME-2: CHiME-1 + More vocab + spk movement (2013) \cite{chime-2} & 6.3,110 & 2 & 9 to -6 & 0.3 & 4,17 @ 6 dB \\
    REVERB: Single speaker in office room (2013) \cite{reverb} & & 1,2.5 & 20 & 0.7 & 30--50 (8--1 mics) \\
    CHiME-3: Tablet with 6 mics outside (2015) \cite{chime-3} & & 0.4 & 5 & 0 & 5.8 \\
    CHiME-4: CHiME 3 + 1,2,6 mics (2016) \cite{chime-4} & & 0.4 & 5 & 0 & 2.2,3.9,9.2 \\
    CHiME-5: Dinner party with multiple talkers (2018) \cite{chime-5} & & 2 & Low & 0.5 & 46 \\
    \hline
    \textbf{Robots at trade shows} & \textbf{Medium/Low} & \textbf{1} & \textbf{15 to 5} & \textbf{1+} & $\mathbf{>50}$ \\
    \hline
    \end{tabular}
    \caption{Factors affecting ASR performance.}
    \label{ASRFactors}
\end{table*}

\section{STATE-OF-THE-ART ASR IN NOISE}
\label{stateOfTheArtASR}

The last section characterized the noisy environment facing public social robots such as XR-1, Pepper, and Cloudia. We now investigate the state-of-the-art available ASR performance in such noisy conditions.

\subsection{Corpus}
\label{corpus}

The AiShell-1 \cite{AiShell} Chinese corpus was used for initial evaluation. AiShell-1 has speech recordings using a high-quality microphone, Android mobile device, and iOS mobile device; we used only the Android recordings since Cloudminds' current robot control unit (RCU), or interface device between robot terminals and a cloud-based platform, is an Android device. Up to now, only the 178-hour open-source part of the AiShell corpus has been used. Up to 718 additional hours of data per recording setup could be acquired if needed.

AiShell-1 comes with pre-partitioned training, development, and test sets of 118664, 14326, and 7176 utterances or 148, 18, and 10 hours, from 336, 40, and 20 speakers. These splits are used for all training and testing.

\subsection{Creating Noisy Data}
\label{creatingNoisyData}

Recorded noise was added to the relatively clean (around 35 dB SNR, as shown in Figure \ref{SNRfigure}) AiShell-1 data to create noisy data. This gives us an immediate large corpus of noisy data which would have been challenging to collect in the field, and also for some techniques such as autoencoders \cite{DAE_Vincent,DAE_Maas} it is necessary to have both clean and corrupted samples from the same data.

We chose SNR increments of 5 dB from 20 dB to 0 dB, and the noise level was scaled to obtain the needed average SNR across a sample of the corpus. Only one noise level was used for a given SNR for the entire corpus, e.g., there is utterance-by-utterance variability in the SNR, but the average is the desired SNR. For a given average SNR condition, the standard deviation of SNR across utterances is 5 dB.

The "base" noise used for adding to the corpus was recorded from Cloudminds' Pepper social robot deployed at the Mobile World Congress trade show in February 2018. We viewed this as more realistic noise than white or pink generated noise. A one minute section of audio was extracted from the raw recording, where there was only background and environmental speech, i.e., no foreground speech. The recordings were made from the front main microphone of a device very much like an Android mobile phone. For each utterance in the training and test corpora, a random piece of that one minute segment was used as the noise portion, to ensure randomness in the added noise.

\subsection{State-of-the-art ASR Engines} \label{sectionComparisonModels}

Today there are a large number of high-performing APIs for ASR from organizations such as Google, iFlyTek, Nuance, Baidu, and Microsoft. These APIs perform very well in relatively clean speech, but we would like to investigate the performance with higher noise that we see with deployments of social robots.

We also explored an open-source Kaldi \cite{Kaldi} Chinese model \cite{CVTE} as another alternative to 3rd-party APIs. This model uses the "chain" variant of the Kaldi TDNN \cite{KaldiChain}, with 40-dimensional filter bank output as features instead of MFCC. Pitch features are not used, and i-vectors are not used for speaker adaptation. The acoustic model was trained on over 2000 hours of speech, and the language model was trained on a 1 TB news corpus.

\subsection{ASR Performance with state-of-the-art engines}

\begin{figure*}[thpb]
      \centering
\begin{tikzpicture}[y=.06cm, x=.5cm,font=\sffamily]
	\draw (0,0) -- coordinate (x axis mid) (25,0);
    \draw (0,0) -- coordinate (y axis mid) (0,90);
    \foreach \x in {25,20,15,10,5}
    	\tikzmath{
        	\y = 25 - \x;
        }
     	\draw (\x,1pt) -- (\x,-3pt)
			node[anchor=north] {\y};
    \foreach \x in {0}
     	\draw (\x,1pt) -- (\x,-3pt)
			node[anchor=north] {clean};
    \foreach \y in {0,10,...,90}
     	\draw (1pt,\y) -- (-3pt,\y) 
     		node[anchor=east] {\y}; 
	\node[below=0.8cm] at (x axis mid) {Average SNR (dB)};
	\node[rotate=90, above=0.8cm] at (y axis mid) {CER (\%)};

    \draw plot[mark=triangle*, mark options={fill=white}] 
		file {Google.data};
    \draw plot[mark=triangle*, mark options={fill=black}] 
		file {Baidu.data};
    \draw plot[mark=square*, mark options={fill=white}] 
		file {iFly.data};
    \draw plot[mark=square*, mark options={fill=black}] 
		file {CVTE.data};
        
    \draw (5,80) rectangle (20,90) node[pos=0.5] {Robot SNR Range};
    \draw[dotted] (0,10) -- (25,10);
    \draw[dashed] (0,15) -- (25,15);
    
	\begin{scope}[shift={(0.5,60)}] 
	\draw[yshift=\baselineskip] (0,0) -- 
		plot[mark=triangle*, mark options={fill=white}] (0.25,0) -- (0.5,0)
		node[right]{Engine 4};
     \draw[yshift=2\baselineskip] (0,0) -- 
		plot[mark=triangle*, mark options={fill=black}] (0.25,0) -- (0.5,0)
		node[right]{Engine 3};
     \draw[yshift=3\baselineskip] (0,0) -- 
		plot[mark=square*, mark options={fill=white}] (0.25,0) -- (0.5,0)
		node[right]{Engine 2};
     \draw[yshift=4\baselineskip] (0,0) -- 
		plot[mark=square*, mark options={fill=black}] (0.25,0) -- (0.5,0)
		node[right]{Engine 1};
	\end{scope}
\end{tikzpicture}
      \caption{Character Error Rate (CER) of various existing systems as a function of SNR.}
      \label{results1}
   \end{figure*}
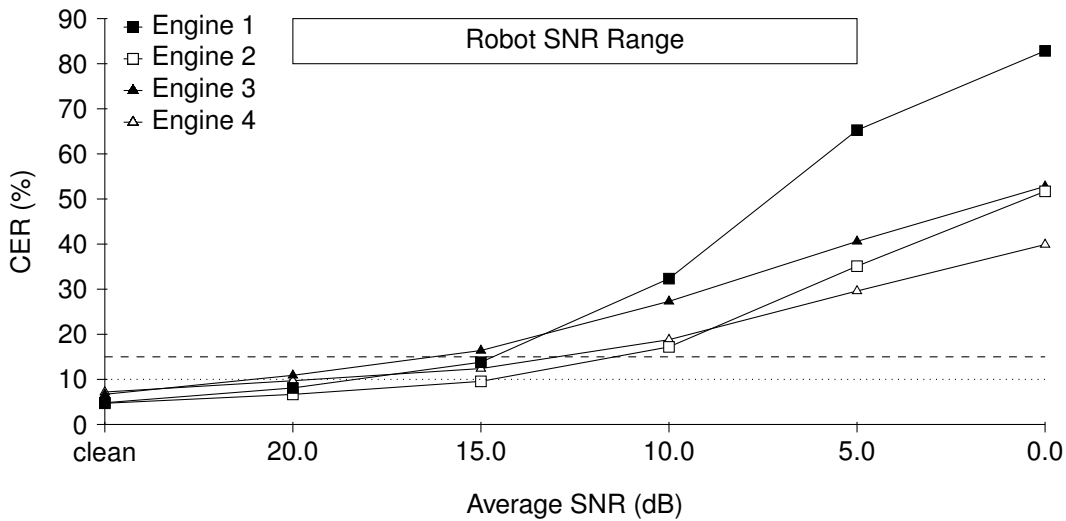

The metric we use for ASR performance is character error rate (CER), which is a standard measure used for Chinese as opposed to word error rate (WER) for many other languages. Figure \ref{results1} shows the CER (using the test set as described in section \ref{corpus}) of the various 3rd-party APIs and open-source models on clean and noisy data. We show the range of SNR for robot deployments, as well as a dotted line at 10\% CER and a dashed line at 15\% CER. For CER exceeding 10-15\%, the usability of system is questionable.

We see that these models perform very well in clean speech and low noise, but CER increases substantially with higher noise, especially at SNR lower than 15 dB. The extent that the model worsens with more noise is quite system-dependent. Given that performance is degrading in the operating region for robots, it is worth investigating methods to reduce error rates for SNR less than 15 dB. 

\section{Robust ASR by training on noisy data}

In previous sections we showed the performance of available APIs and models on noisy speech; this has motivated investigating the extent that ASR models trained or adapted on noisy conditions could improve ASR performance and bring it into an area of reasonable usability.

\subsection{ASR System}

All evaluations use the Kaldi speech recognition toolkit \cite{Kaldi}. 

\subsubsection{Acoustic Model}

We use the AiShell Kaldi recipe \cite{AiShell}, which uses monophone then triphone-based GMM models, using first MFCCs plus deltas and then multiple-frame LDA plus MLLT \cite{MLLT}, then speaker adaptation using fMLLR \cite{fMLLR}, and finally the DNN "chain" model \cite{KaldiChain} incorporating online iVectors \cite{ivectors} for speaker characteristics.

\subsubsection{Language Model}

For fair comparison with 3rd party ASR APIs, which handle a wide variety of text, it was important to use a more general language model than the one trained only on AiShell data, which is the default in the Kaldi recipe. The perplexity of such general language models will be significantly higher, thus resulting in lower ASR accuracy, compared to language models trained only on the ASR training corpus.

We used the Sogou Chinese news corpus \cite{Sogou}, which contained roughly 600 GB of text. A trigram language model was built using Witten-Bell smoothing \cite{WittenBell} and appropriate pruning to be computationally feasible without sacrificing much performance.

\subsubsection{Multi-condition training}

Acoustic models were trained with the original AiShell-1 training set of 148 hours, plus noisy versions of the training set at 20, 15, 10, 5, and 0 dB SNR, for a total training set of 888 hours. 

Training took approximately 1 CPU-week using machines with Nvidia P4 GPUs (only 1 GPU per machine was used even if the machine had more than 1).

\subsection{Results}
\label{ResultsSingleMulti}

\begin{figure*}[thpb]
      \centering
\begin{tikzpicture}[y=.1cm, x=.5cm,font=\sffamily]
	\draw (0,0) -- coordinate (x axis mid) (25,0);
    \draw (0,0) -- coordinate (y axis mid) (0,60);
    \foreach \x in {25,20,15,10,5}
    	\tikzmath{
        	\y = 25 - \x;
        }
     	\draw (\x,1pt) -- (\x,-3pt)
			node[anchor=north] {\y};
    \foreach \x in {0}
     	\draw (\x,1pt) -- (\x,-3pt)
			node[anchor=north] {clean};
    \foreach \y in {0,10,...,60}
     	\draw (1pt,\y) -- (-3pt,\y) 
     		node[anchor=east] {\y}; 
	\node[below=0.8cm] at (x axis mid) {Average SNR (dB)};
	\node[rotate=90, above=0.8cm] at (y axis mid) {CER (\%)};

    \draw (5,54) rectangle (20,60) node[pos=0.5] {Robot SNR Range};
    \draw[dotted] (0,10) -- (25,10);
    \draw[dashed] (0,15) -- (25,15);


    \draw plot[mark=otimes*, mark options={fill=white}] 
		file {kaldi-chain-multi-wwwb-offline.dat};
    \draw plot[mark=*, mark options={fill=black}] 
		file {iFly.data};
    \draw plot[mark=triangle*, mark options={fill=black}] 
		file {Google.data};
    
	\begin{scope}[shift={(1,30)}] 
    \draw[yshift=4\baselineskip] (0,0) -- 
		plot[mark=oplus*, mark options={fill=white}] (0.25,0) -- (0.5,0) 
		node[right]{Kaldi chain, multi-condition noise};
     \draw[yshift=5\baselineskip] (0,0) -- 
		plot[mark=*, mark options={fill=black}] (0.25,0) -- (0.5,0)
		node[right]{Engine 2};
     \draw[yshift=6\baselineskip] (0,0) -- 
		plot[mark=triangle*, mark options={fill=black}] (0.25,0) -- (0.5,0)
		node[right]{Engine 4};
	\end{scope}
\end{tikzpicture}
      \caption{Character Error Rate (CER) as a function of SNR, including custom model.}
      \label{results2}
   \end{figure*}
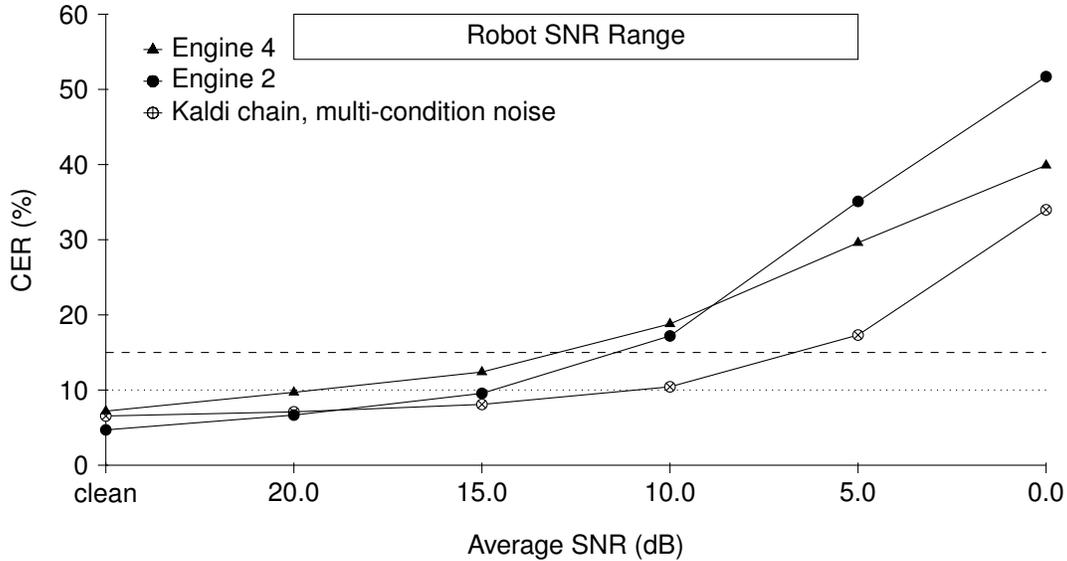

Table \ref{results2table} and figure \ref{results2} show the CER results of the two best performing engines from figure \ref{results1}, as well as the new multi-condition-trained custom models.

Table \ref{results2table} shows the results for engines 2 and 4 from figure \ref{results1}, as well as the binomial standard deviation (using the minimum of engines 2 and 4 as the baseline) and number of standard deviations by which the custom model exceeds the best of engines 2 and 4. Noise conditions where the custom model exceed performance by more than 4 standard deviations are shown in bold.

\begin{table}
    \centering
    \begin{tabular}{|l|l|l|l|l|l|l|}
    \hline
    ASR & Clean & 20 dB & 15 dB & 10 dB & 5 dB & 0 dB \\
    \hline
    Eng2 & 4.7 & 6.7 & 9.6 & 17 & 35 & 52 \\
    Eng4 & 7.2 & 9.7 & 12 & 19 & 30 & 40 \\
    \hline
    Custom & 6.6 & 7.1 & \textbf{8.1} & \textbf{10} & \textbf{17} & \textbf{34} \\
    \hline
    sdev & 0.25 & 0.30 & 0.35 & 0.44 & 0.54 &0.58 \\
    \#sdev & -7.6 & -1.4 & \textbf{4.3} & \textbf{16} & \textbf{24} & \textbf{10} \\
    \hline
    \end{tabular}
    \caption{Character Error Rate (CER) as a function of SNR, including custom model.}
    \label{results2table}
\end{table}

We see that for SNRs of 15 dB or less, the custom-trained models performed statistically significantly better than the best of the existing engines. At 20 dB SNR the difference in results are not significant. For clean speech, the existing engines do significantly better, which is expected given the large amount of time and data behind these models.

\section{Implementation}

Since the goal of this work is to improve real-world performance of ASR on actual deployed robots, it is imperative to operate with a framework that would readily facilitate deployment of the custom-trained models.

We used the gStreamer interface to Kaldi \cite{KaldiGstreamer,KaldiGstreamerWebsite} and the associated Docker image for this interface \cite{KaldiGstreamerDocker} to quickly deploy the Kaldi-trained model.

We have used this infrastructure to deploy the custom models on various social robots in the field. One of many advantages of using the Kaldi toolkit is the small amount of effort needed for deployment, so activities can be focused on model development.

\section{FUTURE WORK}

We plan on exploring many directions in order to continue to improve ASR performance:

\begin{itemize}

\item Other languages: This work is so far only in Chinese. We plan to extend to English with the Librispeech corpus \cite{librispeech} and Japanese with the CSJ corpus \cite{CSJ}, both which have established Kaldi recipes.

\item Additional Data: Figure \ref{resultsTrainingData} shows that there is still considerable room for performance improvement, even in clean conditions. One next step would thus be to add additional training data. For example, we could incorporate the 1000 hour AiShell-2 corpus \cite{AiShell-2}.

\begin{figure}[thpb]
      \centering
\begin{tikzpicture}[y=.1cm, x=.03cm,font=\sffamily]
	\draw (0,0) -- coordinate (x axis mid) (150,0);
    \draw (0,0) -- coordinate (y axis mid) (0,30);
    \foreach \x in {0,50,...,150}
     	\draw (\x,1pt) -- (\x,-3pt)
			node[anchor=north] {\x};
    \foreach \y in {0,10,...,30}
     	\draw (1pt,\y) -- (-3pt,\y) 
     		node[anchor=east] {\y}; 
	\node[below=0.8cm] at (x axis mid) {Training data (hours)};
	\node[rotate=90, above=0.8cm] at (y axis mid) {CER (\%)};

	\draw plot[mark=*, mark options={fill=white}] 
		file {kaldi-chain-clean.dat};
    \draw plot[mark=triangle*, mark options={fill=white}] 
		file {kaldi-chain-MWC20.dat};
    \draw plot[mark=triangle*, mark options={fill=black}] 
		file {kaldi-chain-MWC15.dat};
    \draw plot[mark=square*, mark options={fill=black}] 
		file {kaldi-chain-MWC5.dat};
    
	\begin{scope}[shift={(120,20)}] 
	
	\draw[yshift=\baselineskip] (0,0) -- 
		plot[mark=square*, mark options={fill=black}] (0.25,0) -- (0.5,0)
		node[right]{5 dB SNR};
	\draw[yshift=4\baselineskip] (0,0) -- 
		plot[mark=*, mark options={fill=white}] (0.25,0) -- (0.5,0)
		node[right]{clean};
     \draw[yshift=2\baselineskip] (0,0) -- 
		plot[mark=triangle*, mark options={fill=black}] (0.25,0) -- (0.5,0)
		node[right]{15 dB SNR};
     \draw[yshift=3\baselineskip] (0,0) -- 
		plot[mark=triangle*, mark options={fill=white}] (0.25,0) -- (0.5,0)
		node[right]{20 dB SNR};
     
	\end{scope}
\end{tikzpicture}
      \caption{CER as a function of hours of training data, for various noise conditions.}
      \label{resultsTrainingData}
   \end{figure}
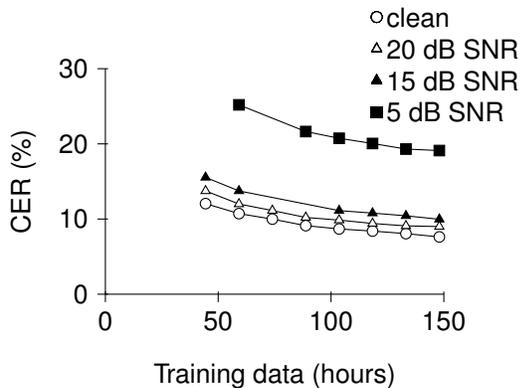

\item Microphone arrays: Especially for social robots in public spaces with multiple possible speakers, sometime simultaneous, it would be useful to use multiple microphones to detect the live speaker and also focus attention on that speaker, decreasing the effective SNR as compared with single microphones. Results from the CHiME challenges using techniques such as Beamformit \cite{beamformit} have been very successful.

\item Dereverberation: Likewise, especially since we saw in table \ref{ASRFactors} that $T_{60}$ for robots can be much worse than other environments, explicit dereverberation techniques such as Weighted Prediction Error (WPE) \cite{WPE} have been very successful in use cases such as the REVERB challenge.

\item Audio-visual speech recognition: Research has shown that, especially for SNR less than 20 dB, Audio-visual speech recognition or "lipreading" can improve ASR performance, sometimes dramatically \cite{AVSR1,AVSR2,AVSR3,CompVisemeReco,MultimodalTaxonomy,CrossDBAVSR}. Since robots such as Pepper will almost invariably have access to still pictures and video, this is a potentially useful technique not available to all ASR deployments such as voice-only bots.
\end{itemize}

\section{Conclusion}

We have shown that the ASR environment for social robots deployed in public spaces is somewhat challenging, and that established ASR models that perform well in clean speech can degrade significantly under such conditions. We were able to train custom ASR models using the Kaldi toolkit that performed better than existing models for SNR less than 20 dB. We have been able to successfully deploy these custom models on robots in the field. We will continue along several paths to improve robust ASR performance for social robots.

This work demonstrates the feasibility of using open-source toolkits to achieve high performing ASR for particularly challenging conditions without enormous amounts of training data.

\bibliography{main}

\begin{thebibliography}{10}
\providecommand{\url}[1]{#1}
\csname url@samestyle\endcsname
\providecommand{\newblock}{\relax}
\providecommand{\bibinfo}[2]{#2}
\providecommand{\BIBentrySTDinterwordspacing}{\spaceskip=0pt\relax}
\providecommand{\BIBentryALTinterwordstretchfactor}{4}
\providecommand{\BIBentryALTinterwordspacing}{\spaceskip=\fontdimen2\font plus
\BIBentryALTinterwordstretchfactor\fontdimen3\font minus
  \fontdimen4\font\relax}
\providecommand{\BIBforeignlanguage}[2]{{%
\expandafter\ifx\csname l@#1\endcsname\relax
\typeout{** WARNING: IEEEtran.bst: No hyphenation pattern has been}%
\typeout{** loaded for the language `#1'. Using the pattern for}%
\typeout{** the default language instead.}%
\else
\language=\csname l@#1\endcsname
\fi
#2}}
\providecommand{\BIBdecl}{\relax}
\BIBdecl

\bibitem{FTC2019}
V.~Mruthyunjaya and C.~Jankowski, ``Human-augmented robotic intelligence (hari)
  for human-robot interaction,'' in \emph{2019 Future Technology Conference},
  2019.

\bibitem{SNRenvironment}
K.~Pearsons, R.~Bennett, and S.~Fidell, ``Speech levels in various noise
  environments,'' U.S. Environmental Protection Agency, Washington, D.C., Tech.
  Rep. EPA/600/1-77/025 (NTIS PB270053), 1977.

\bibitem{GoogleHomeSNR}
T.~Sainath, Informal communication, December 2017, aSRU 2017.

\bibitem{chime-1}
J.~Barker, E.~Vincent, N.~Ma, H.~Christensen, and P.~Green, ``The {PASCAL}
  {CHiME} speech separation and recognition challenge,'' \emph{Computer Speech
  and Language}, vol.~27, no.~3, pp. 621--633, May 2013.

\bibitem{chime-2}
E.~Vincent, J.~Barker, S.~Watanabe, J.~Le~Roux, F.~Nesta, and M.~Matassoni,
  ``The second `chime’ speech separation and recognition challenge: Datasets,
  tasks and baselines.'' in \emph{Proc. IEEE International Conference on
  Acoustics, Speech and Signal Processing}, 2013.

\bibitem{chime-3}
E.~V. Jon~Barker, Ricard~Marxer and S.~Watanabe, ``The third ‘chime’ speech
  separation and recognition challenge: Analysis and outcomes,'' \emph{Computer
  Speech and Language}, vol.~46, pp. 605--626, 2017.

\bibitem{chime-4}
E.~Vincent, S.~Watanabe, A.~Nugraha, J.~Barker, and R.~Marxer, ``An analysis of
  environment, microphone and data simulation mismatches in robust speech
  recognition,'' \emph{Computer Speech and Language}, vol.~46, pp. 535--557,
  2017.

\bibitem{chime-5}
J.~Barker, S.~Watanabe, E.~Vincent, and J.~Trmal, ``The fifth 'chime' speech
  separation and recognition challenge: Dataset, task and baselines,'' 03 2018.

\bibitem{reverb}
K.~Kinoshita, M.~Delcroix, T.~Yoshioka, T.~Nakatani, E.~Habets, R.~Haeb-Umbach,
  V.~Leutnant, A.~Sehr, W.~Kellermann, R.~Maas, S.~Gannot, and B.~Raj, ``The
  reverb challenge: A common evaluation framework for dereverberation and
  recognition of reverberant speech,'' 10 2013.

\bibitem{AiShell}
\BIBentryALTinterwordspacing
H.~Bu, J.~Du, X.~Na, B.~Wu, and H.~Zheng, ``{AISHELL-1:} an open-source
  mandarin speech corpus and {A} speech recognition baseline,'' \emph{CoRR},
  vol. abs/1709.05522, 2017. [Online]. Available:
  \url{http://arxiv.org/abs/1709.05522}
\BIBentrySTDinterwordspacing

\bibitem{DAE_Vincent}
\BIBentryALTinterwordspacing
P.~Vincent, H.~Larochelle, Y.~Bengio, and P.-A. Manzagol, ``Extracting and
  composing robust features with denoising autoencoders,'' in \emph{Proceedings
  of the 25th International Conference on Machine Learning}, ser. ICML
  '08.\hskip 1em plus 0.5em minus 0.4em\relax New York, NY, USA: ACM, 2008, pp.
  1096--1103. [Online]. Available:
  \url{http://doi.acm.org/10.1145/1390156.1390294}
\BIBentrySTDinterwordspacing

\bibitem{DAE_Maas}
A.~L. Maas, Q.~V. Le, T.~M. O'Neil, O.~Vinyals, P.~Nguyen, and A.~Y. Ng,
  ``Recurrent neural networks for noise reduction in robust asr,'' in
  \emph{INTERSPEECH 2012, 13th Annual Conference of the International Speech
  Communication Association, Portland, Oregon, USA, September 9-13,
  2012}.\hskip 1em plus 0.5em minus 0.4em\relax ISCA, 2012, pp. 22--25.

\bibitem{Kaldi}
D.~Povey, A.~Ghoshal, G.~Boulianne, L.~Burget, O.~Glembek, N.~Goel,
  M.~Hannemann, P.~Motlicek, Y.~Qian, P.~Schwarz, J.~Silovsky, G.~Stemmer, and
  K.~Vesely, ``The kaldi speech recognition toolkit,'' 2011, iEEE Catalog No.:
  CFP11SRW-USB.

\bibitem{CVTE}
\BIBentryALTinterwordspacing
D.~Povey. Cvte mandarin model. [Online]. Available:
  \url{http://kaldi-asr.org/models/m2}
\BIBentrySTDinterwordspacing

\bibitem{KaldiChain}
V.~Peddinti, D.~Povey, and S.~Khudanpur, ``A time delay neural network
  architecture for efficient modeling of long temporal contexts,'' in
  \emph{INTERSPEECH}, 2015.

\bibitem{MLLT}
R.~A. Gopinath, ``Maximum likelihood modeling with gaussian distributions for
  classification,'' in \emph{Acoustics, Speech and Signal Processing, 1998.
  Proceedings of the 1998 IEEE International Conference on}, vol.~2, May 1998,
  pp. 661--664 vol.2.

\bibitem{fMLLR}
M.~Gales, ``Maximum likelihood linear transformations for hmm-based speech
  recognition,'' \emph{Computer Speech \& Language}, vol.~12, no.~2, pp. 75 --
  98, 1998.

\bibitem{ivectors}
N.~Dehak, P.~J. Kenny, R.~Dehak, P.~Dumouchel, and P.~Ouellet, ``Front-end
  factor analysis for speaker verification,'' \emph{IEEE Transactions on Audio,
  Speech, and Language Processing}, vol.~19, no.~4, pp. 788--798, May 2011.

\bibitem{Sogou}
\BIBentryALTinterwordspacing
C.~Wang, M.~Zhang, S.~Ma, and L.~Ru, ``Automatic online news issue construction
  in web environment,'' in \emph{Proceedings of the 17th International
  Conference on World Wide Web}, ser. WWW '08.\hskip 1em plus 0.5em minus
  0.4em\relax New York, NY, USA: ACM, 2008, pp. 457--466. [Online]. Available:
  \url{http://doi.acm.org/10.1145/1367497.1367560}
\BIBentrySTDinterwordspacing

\bibitem{WittenBell}
I.~Witten and T.~Bell, ``The zero-frequency problem: Estimating the
  probabilities of novel events in adaptive text compression,'' \emph{IEEE
  Transactions on Information Theory}, vol.~37, pp. 1085--1094, 07 1991.

\bibitem{KaldiGstreamer}
T.~Alum\"{a}e, ``Full-duplex speech-to-text system for {Estonian},'' in
  \emph{Baltic HLT 2014}, Kaunas, Lithuania, 2014.

\bibitem{KaldiGstreamerWebsite}
\BIBentryALTinterwordspacing
------. Real-time full-duplex speech recognition server, based on the kaldi
  toolkit and the gstreamer framwork. [Online]. Available:
  \url{https://github.com/alumae/kaldi-gstreamer-server}
\BIBentrySTDinterwordspacing

\bibitem{KaldiGstreamerDocker}
\BIBentryALTinterwordspacing
E.~Silva. Dockerfile for kaldi-gstreamer-server. [Online]. Available:
  \url{https://github.com/jcsilva/docker-kaldi-gstreamer-server}
\BIBentrySTDinterwordspacing

\bibitem{librispeech}
V.~Panayotov, G.~Chen, D.~Povey, and S.~Khudanpur, ``Librispeech: An asr corpus
  based on public domain audio books,'' in \emph{2015 IEEE International
  Conference on Acoustics, Speech and Signal Processing (ICASSP)}, April 2015,
  pp. 5206--5210.

\bibitem{CSJ}
\BIBentryALTinterwordspacing
K.~Maekawa, ``Corpus of spontaneous japanese : Its design and evaluation,''
  \emph{Proc. ISCA \& IEEE Workshop SSPR 2003}, 2003. [Online]. Available:
  \url{https://ci.nii.ac.jp/naid/10021838110/en/}
\BIBentrySTDinterwordspacing

\bibitem{AiShell-2}
\BIBentryALTinterwordspacing
J.~Du, X.~Na, X.~Liu, and H.~Bu, ``{AISHELL-2:} transforming mandarin {ASR}
  research into industrial scale,'' \emph{CoRR}, vol. abs/1808.10583, 2018.
  [Online]. Available: \url{http://arxiv.org/abs/1808.10583}
\BIBentrySTDinterwordspacing

\bibitem{beamformit}
X.~A. Mir{\'o}, C.~Wooters, and J.~Hernando, ``Acoustic beamforming for speaker
  diarization of meetings,'' \emph{IEEE Transactions on Audio, Speech, and
  Language Processing}, vol.~15, pp. 2011--2022, 2007.

\bibitem{WPE}
T.~{Nakatani}, T.~{Yoshioka}, K.~{Kinoshita}, M.~{Miyoshi}, and B.~{Juang},
  ``Speech dereverberation based on variance-normalized delayed linear
  prediction,'' \emph{IEEE Transactions on Audio, Speech, and Language
  Processing}, vol.~18, no.~7, pp. 1717--1731, Sep. 2010.

\bibitem{AVSR1}
S.~Lucey, S.~Sridharan, and V.~Chandran, ``Improved speech recognition using
  adaptive audio-visual fusion via a stochastic secondary classifier,'' in
  \emph{Proceedings of 2001 International Symposium on Intelligent Multimedia,
  Video and Speech Processing. ISIMP 2001 (IEEE Cat. No.01EX489)}, 2001, pp.
  551--554.

\bibitem{AVSR2}
\BIBentryALTinterwordspacing
J.~Ngiam, A.~Khosla, M.~Kim, J.~Nam, H.~Lee, and A.~Y. Ng, ``Multimodal deep
  learning,'' in \emph{Proceedings of the 28th International Conference on
  International Conference on Machine Learning}, ser. ICML'11.\hskip 1em plus
  0.5em minus 0.4em\relax USA: Omnipress, 2011, pp. 689--696. [Online].
  Available: \url{http://dl.acm.org/citation.cfm?id=3104482.3104569}
\BIBentrySTDinterwordspacing

\bibitem{AVSR3}
\BIBentryALTinterwordspacing
K.~Noda, Y.~Yamaguchi, K.~Nakadai, H.~G. Okuno, and T.~Ogata, ``Audio-visual
  speech recognition using deep learning,'' \emph{Applied Intelligence},
  vol.~42, no.~4, pp. 722--737, Jun 2015. [Online]. Available:
  \url{https://doi.org/10.1007/s10489-014-0629-7}
\BIBentrySTDinterwordspacing

\bibitem{CompVisemeReco}
\BIBentryALTinterwordspacing
D.~Jachimski, A.~Czyzewski, and T.~Ciszewski, ``A comparative study of english
  viseme recognition methods and algorithms,'' \emph{Multimedia Tools and
  Applications}, vol.~77, no.~13, pp. 16\,495--16\,532, Jul 2018. [Online].
  Available: \url{https://doi.org/10.1007/s11042-017-5217-5}
\BIBentrySTDinterwordspacing

\bibitem{MultimodalTaxonomy}
\BIBentryALTinterwordspacing
T.~Baltrusaitis, C.~Ahuja, and L.~Morency, ``Multimodal machine learning: {A}
  survey and taxonomy,'' \emph{CoRR}, vol. abs/1705.09406, 2017. [Online].
  Available: \url{http://arxiv.org/abs/1705.09406}
\BIBentrySTDinterwordspacing

\bibitem{CrossDBAVSR}
S.~Kalantari, D.~Dean, H.~Ghaemmaghami, S.~Sridharan, and C.~Fookes, ``Cross
  database training of audio-visual hidden markov models for phone
  recognition,'' in \emph{INTERSPEECH}, 2015.

\end{thebibliography}
\bibliographystyle{IEEEtrans}

\end{document}